\documentclass[doublecol]{epl2} 

\usepackage{footmisc,multirow,amsmath,amssymb}
\usepackage{subfigure}
\usepackage{amsmath,slashed,enumerate}

\usepackage{graphicx}
\usepackage[normalem]{ulem}

\renewcommand{\d}{{\text{d}}}
\newcommand{\ee}{\end{eqnarray*}}

\newcommand{\bee}{\begin{eqnarray}}
\newcommand{\eee}{\end{eqnarray}}
\newcommand{\beeq}{\begin{equation}}
\newcommand{\eeeq}{\end{equation}}

\renewcommand{\vec}{\bf}

\title{Particle Physics with Gravitational Wave Detector Technology}

\author{
Christoph~Englert\inst{1}
\and
Stefan~Hild\inst{1}
\and
Michael~Spannowsky\inst{2}
}

\institute{   
\inst{1} SUPA, School of Physics and Astronomy, University of Glasgow, Glasgow G12 8QQ, UK\\
\inst{2} Institute for Particle Physics Phenomenology, Department of Physics, Durham University, Durham DH1 3LE, UK
}
 
 
\abstract{
Gravitational wave detector technology provides high-precision measurement apparatuses that, if combined with a modulated particle source,
have the potential to measure and constrain particle interactions in a novel way, by measuring
the pressure caused by scattering particle beams off the mirror material. Such a measurement does not rely on tagging a final state.
This strategy
has the potential to allow us to explore novel ways to constrain the presence of new interactions beyond the Standard Model of Particle 
Physics and provide additional constraints to poorly understood cross sections in the non-perturbative regime of 
QCD and Nuclear Physics, which are limiting factors of dark matter and neutrino physics searches. Beyond high-energy physics, if technically feasible, the proposed method to measure nucleon-nucleon interactions can lead to practical applications in material and medical sciences.
}

\begin{document}

\maketitle

%
%
\section{Introduction}
\label{sec:intro}
The direct detection of gravitational waves~\cite{Abbott:2016blz,Abbott:2016nmj,Abbott:2017oio,TheLIGOScientific:2017qsa} marks the beginning of a new era of Astronomy, Cosmology and Astrophysics that will exhaust the opportunities introduced by high precision interferometry techniques developed for gravitational wave detectors such as LIGO. The precision with which measurements can be performed opens up the prospects of better understanding early Universe phenomena such as baryogenesis~\cite{Grojean:2006bp}, exotic physics on cosmological scales~\cite{Caprini:2015zlo,Jaeckel:2016jlh,Schwaller:2015tja}, test the nature of gravity~\cite{Cardoso:2016rao,Abedi:2016hgu,Baker:2017hug,Sakstein:2017xjx,Ezquiaga:2017ekz} and constrain aspects of the cosmological standard model~\cite{Schutz:1986gp,Holz:2005df,Creminelli:2017sry} through their gravitational signals.

Most of the implications of gravitational wave observations evolve around the classical features of gravity as well as its potential modifications. However, the high precision that is offered by gravitational wave detectors and their underlying working principles could offer new opportunities for particle physics as new sensitive probes of particle interactions. Especially in the low energy limit of Quantum Chromodynamics, hadronic cross sections are plagued by big theoretical as well as experimental uncertainties, that feed into a series of searches for beyond the Standard Model interactions. For instance, hadronic and nuclear interactions are key limiting factors for searches for new effects in the neutrino sector~\cite{Katori:2016yel}, where additional information could be used to gain a more fine-grained picture of multi-nucleon interaction and nucleon correlation~\cite{CiofidegliAtti:2017xtx}. 

Using the sensitivity of gravitational wave detectors' mirrors to smallest forces, a gravitational wave detector, or smaller interferometer providing a similar force sensitivity, but without the need to build kilometre-scale arms, can in principle be turned into a particle physics detector through measuring the pressure caused by  scattering of a (modulated) beam off the material. Such a measurement can be, but does not have to be correlated with observation of transmission. Inclusive scattering cross sections can therefore be measured without relying on final state particle information if a certain material is sufficiently well-understood.

After discussing the sensitivity provided by gravitational wave detector technology, we calculate the expected pressures in a range of simplified scenarios that allow us to correlate pressure and total scattering cross section straightforwardly. Considering realistic estimates of beam conditions of sources of highest intensity, we  argue on theoretical grounds that the expected sensitivity is high enough to access strong interaction cross sections in a completely novel way.

\begin{figure*}[!t]
	\subfigure[\label{fig:SSM}~Target displacement sensitivity.]{\includegraphics[width=8.5cm]{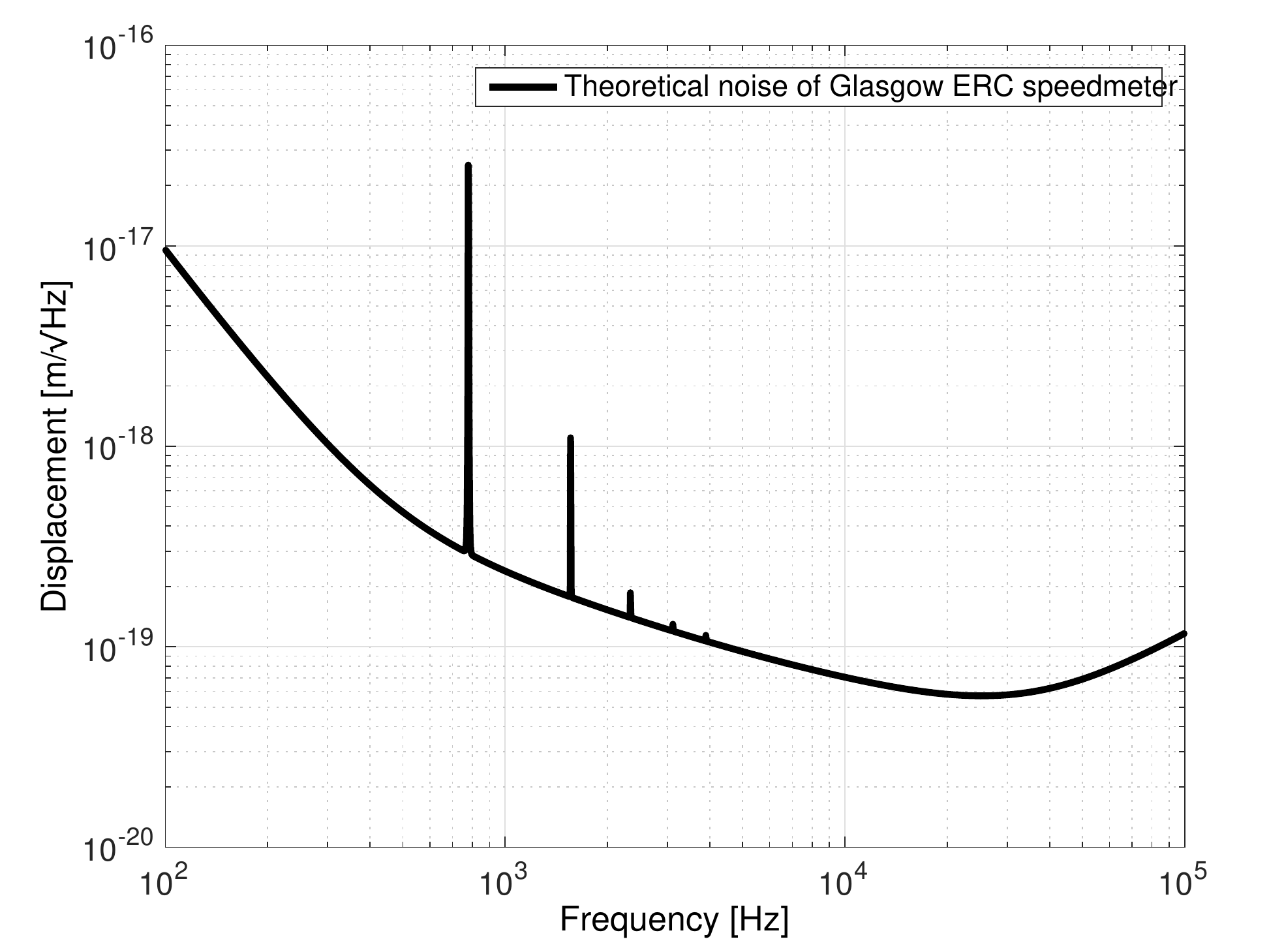}}\hfill
	\subfigure[\label{fig:Force} Target force sensitivity.]{\includegraphics[width=8.5cm]{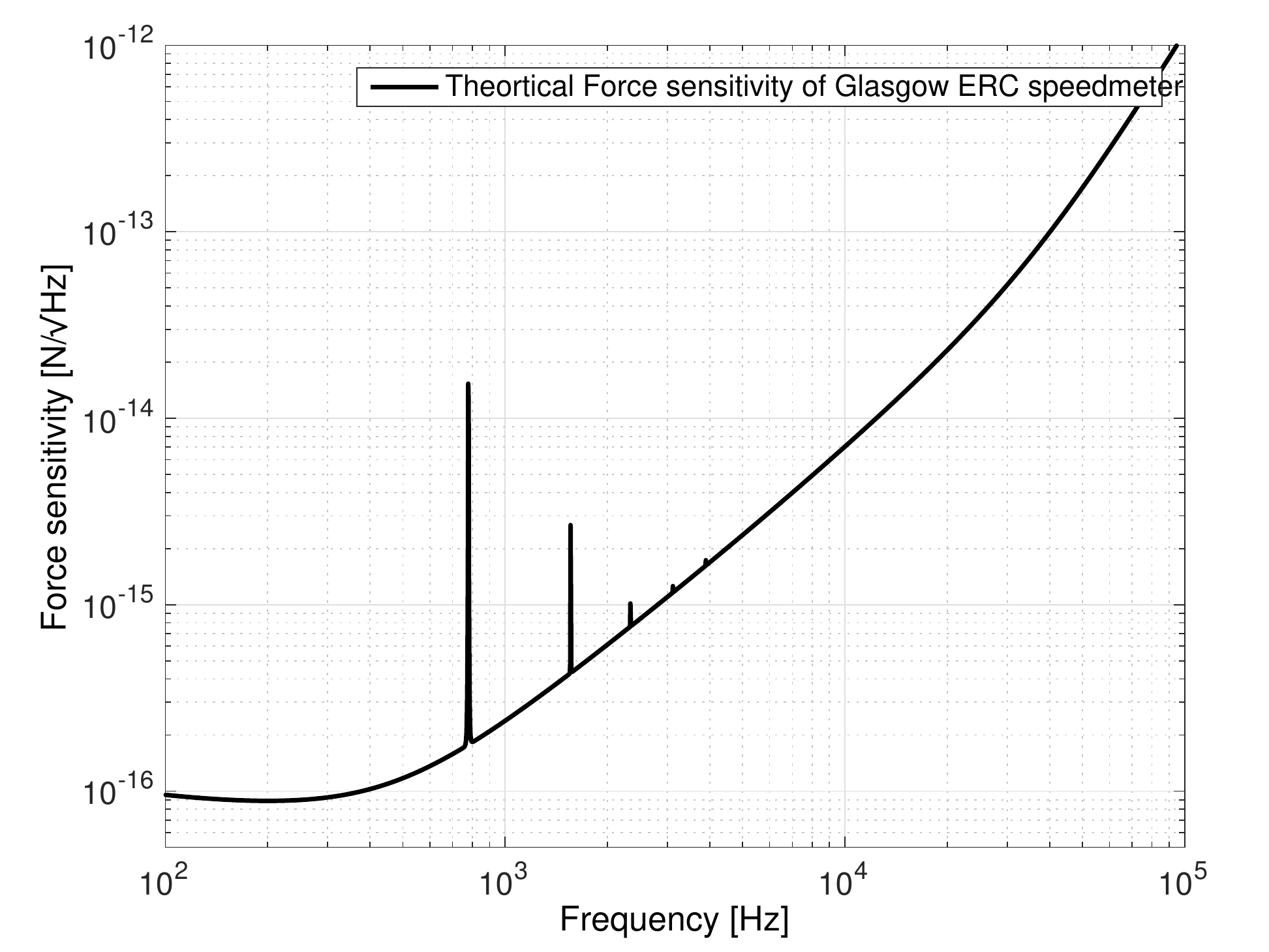}}
	\caption{\label{fig:intens}Sensitivity of the Glasgow ERC speedmeter experiment. The resonances with decreased sensitivity are related to the violin modes of the mirror suspension fibres.}
\end{figure*}

\section{Achievable Force Sensitivity Using Gravitational Wave Detector Technologies}

Gravitational Wave Detectors such as LIGO, GEO~600 or VIRGO
 \cite{LIGO, GEO, Virgo}  have 
established high-precision measurements of differential displacement of end mirrors of their 
orthogonal arms, reaching sensitivities in the range of $10^{-19}$~m/$\sqrt{\rm Hz}$ for 
frequencies roughly covering the audio-band. Employing technology similar to large-scale
laser-interferometric GW detectors, one can conceive meter-scale experiments\footnote{Note
that the kilometre-scale arm length for real GW detectors is required to increase its 
response to spacetime fluctuations. However, for simple force measurement the 
achievable sensitivity will be to first order independent of the length of the laser interferometer.} which can provide a similar displacement sensitivity in the kHz range.
In the following we will use the example of the Glasgow ERC speedmeter interferometer
\cite{SSM-design} to show 
what range of forces applied to one of the mirrors of the interferometer can be measured. 
The speedmeter configuration suggested
here provides the advantage of cancelling some
of the quantum back action noise and therefore
it provides a better force sensitivity than scaled versions of LIGO or Virgo. For details see~\cite{SSM-design}. 
We chose that experiment because of its small scale  in terms of cost (about 1 million~\$) and space (footprint of about $4\times 2$~m), which makes
it conceivable to consider to set up a copy of the experiment close to the beamline of a particle 
accelerator. 

Interferometry-based approaches to particle physics measurements are not new and have been discussed in particular in the context of neutrino physics~\cite{Shvartsman:1982sn,Domcke:2017aqj}. The crucial difference compared to our setup is the controlled modulation of the beam which places signals into a frequency range that is accessible by terrestrial experiments with high precision.

\begin{figure}[!b]
	\includegraphics[width=8.5cm]{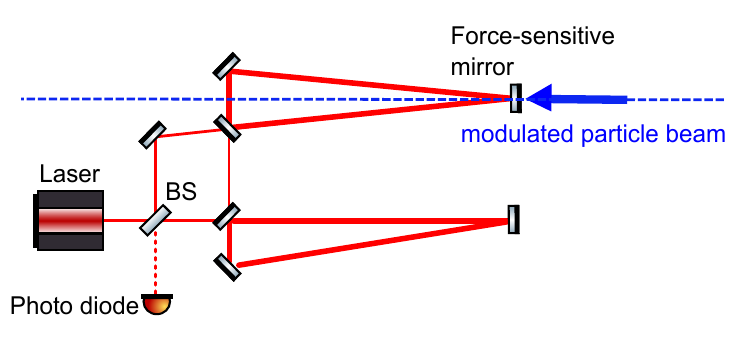}
	\caption{\label{fig:setup} Schematic layout of the proposed experimental setup. Red lines indicate the laser beams of reading out the differential length of the two triangular arm cavities. One of the cavity end mirrors is used as the target of a modulated particle 
	beam indicated by the blue arrow. }
\end{figure}

Figure~\ref{fig:SSM} shows the design sensitivity of the Glasgow ERC speedmeter, 
expressed as the linear spectral density of differential displacements of its interferometer
mirrors. Each of the arm cavity resonators of the laser interferometer features a mirror of
mass $m = 1$~g, 
with a diameter of 10~mm, suspended from a multi-stage pendulum. In the 
following sections we assume that a particle beam, fully modulated at a frequency $f$ 
is focussed down to less than 10~mm in diameter and impinges onto one of the interferometer 
mirrors. We also assume that apart from this probing mirror, no other component of the
 laser interferometer is influenced by the modulated particle beam or the apparatus 
 creating it. For a sketch of the experimental setup see Fig.~\ref{fig:setup}. Then we can simply compute the linear spectral density of the force sensitivity of the ERC speedmeter for forces applied to one of its 1 gram mirrors:
 \begin{equation}
 F(f) =  X_{sens}(f) m f^2 , 
 \end{equation}
 where ${X_{sens}}$ is the equivalent displacement spectral density shown in Fig.~\ref{fig:SSM}.
The resulting spectral density of the force sensitivity, given in units of Newton per 
square root of Hz, is displayed in Fig.~\ref{fig:SSM}. For a particle beam 
modulation frequency of 1~kHz a force of $2\times 10^{-16}$~N exerted onto the mirror 
would be measured with a signal to noise ratio of 1 for a measurement of duration 1 second.
This sensitivity is limited by noise processes inherent to the Glasgow ERC speedmeter
such Brownian fluctuations of the molecules in the mirror coating and mirror 
suspension fibres \cite{Pitkin2011} or quantum noise, a combination of
 sensing and back-action noise of the 
photons in the interferometer \cite{Danilishin2012}. However, obviously 
the achievable signal to noise ratio 
can easily be increased by lengthening the measurement duration. Similarly, integration 
over longer durations allows the observation of forces smaller than the level shown 
Fig.~\ref{fig:SSM}. In the following sections we will conservatively focus on the sensitivity achievable with a measuring duration t of 1 second. However, we point out that the sensitivity improves with $\sqrt{\text{time}}$, i.e., for $t=100~\text{s}$ the sensitivity improves by a factor of 10 and for one week of data-taking it improves by a factor of $\sim 777$.

\section{Pressure from scattering}
Macroscopic pressure can be related to scattering through the momentum transfer between incident beam and target material per unit area and unit of time. In the microscopic picture these effects are related to $2\to 2$ scattering processes with differential cross section $\d\sigma$. Choosing the beam axis in $z$ direction the pressure is evaluated by weighting the momentum transfer in $z$ direction with the corresponding field theoretic probability for a simplified geometry (see e.g.~\cite{Adler:1974rj})
\begin{equation}
\label{eq:cohscatt}
P\simeq{\cal{F}} {\cal{T}} \int_{-1}^{1} \d \alpha \, {\d \sigma \over \d \alpha} \, p_z \, (1-\alpha)\,.
\end{equation}
$p_z$ is the $z$-component of the incident particle's momentum that is reduced by a factor $\alpha\in[-1,1]$ by scattering off the target material. ${\cal F}$ denotes the flux of incoming particles per unit area and time and $\cal{T}$ is the optical thickness measured in number per unit area. The flux can be controlled in the experimental setting while $\cal{T}$ is a material-dependent quantity. 

In particle and nuclear physics-based collider experiments such as the Large Hadron Collider, total (or exclusive) cross sections are also inferred from an underlying differential cross section
\begin{equation}
\label{eq:interaction}
\sigma = \int_{-1}^{1} \d \alpha \, {\d \sigma \over \d \alpha}\,.
\end{equation}
We can therefore correlate event count measurements of scattering processes at colliders with pressure constraints for a given theory model that underpins $\d \sigma$. As the flux can be controlled experimentally, the implications are two-fold: if we have a good understanding of the scattering cross section, the material-dependent parameter $\cal{T}$ can be inferred. If $\cal{T}$ is sufficiently known, Eq.~\eqref{eq:cohscatt} provides a complementary constraint on our modelling of $\d \sigma$. A more detailed modelling of the beam-absorber interaction can be achieved efficiently using GEANT \cite{Agostinelli:2002hh}. Software frameworks like GEANT allow the inclusion of multiple incident particle-material scatterings, ionisation effects etc; radiation-induced detector degrading can be included as well. Although these processes are all relevant, we neglect them in the following to highlight different pressure-model correlations. Furthermore, changing the beam particles from protons to weakly interacting particles, e.g. neutrinos or muons, secondary interactions with the absorber material and systematic uncertainties would be significantly modified. 
Mirror deformations can in principle occur if the
forces acting on the mirror are not homogeneously
distributed. However as shown in Ref.~\cite{Hild:2007rr} such effects only become important at frequencies
above the first body resonances. For the mirror dimensions
suggested in this article this would be several tens
of kHz, and hence far away from the suggested measurement
frequency. Therefore, we do not expect this effect
to cause any measurable effect. Our results below should  be understood as proof-of-concept rather than a precision study for an existing experimental setup.

\begin{figure}[!t]
	\centering
	\includegraphics[width=3.3cm]{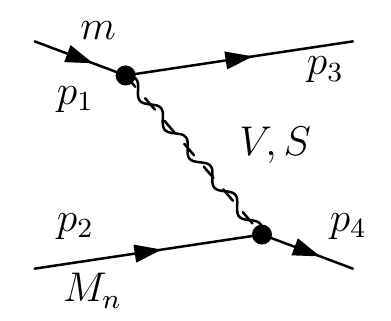}
	\caption{\label{fig:feyn} Scattering in the toy models discussed in this work. The scalar $(S)$ and vectorial mediators $(V)$ carry momentum
	transfer $t=(p_1-p_3)^2=(p_2-p_4)^2$ with four momenta $p_1^2,p_3^2=m^2$ and $p_2^2,p_4^2=M_n^2$ in the elastic case. The interaction vertices are generated
	by the interactions quoted in Eq.~\eqref{eq:interaction}.}
\end{figure}

We will consider elastic scattering in the following: $p_1(m) +p_2(M_n) \to p_3(m) + p_4(M_n)$, with $m$ denoting the mass of the incident particle and $M_n$ the target mass (we will comment on inelastic scattering below). We focus on $t$-channel mediators, $t=(p_1-p_3)^2<0$ of the scattering and consider scalar and vectorial toy interactions with different Lorentz structures
\begin{equation}
\label{eq:lagrange}
{\cal{L}}=\sum_i \bar\Psi_i (c_1 S + c_2 \gamma^\mu V_\mu )
\Psi_i
\end{equation}
to highlight complementarity of the pressure measurement for a given cross section value (see Fig.~\ref{fig:feyn}). We denote the 
mediator masses with $m_{S,V}$, respectively.
The sum runs over our mass choices $i=m,M_n$. The effective couplings $c_i$ and masses are model-dependent and can have momentum transfer-dependencies. For instance, search strategies for dark matter in the context of simplified models do typically neglect any momentum dependencies in first instance~(e.g.~\cite{DiFranzo:2013vra}). We will choose $c_i$ as constants to highlight the different pressure-cross section correlations most transparently.

Equation~\eqref{eq:cohscatt} (including the momentum transfer of the scattering) can be obtained from the amplitude ${\cal{M}}$ for fixed target kinematics via~(see e.g.~\cite{Patrignani:2016xqp})
\begin{equation}
\label{eq:alpha}
{\d\sigma \over \d \alpha}= {1\over 64\pi s} {1 \over |{\vec{p}}_{1,{\text{cm}}}|^2}  |{\cal{M}}|^2  {\d t\over \d \alpha}\,,
\end{equation}
with $s=(p_1+p_2)^2$ and
\begin{equation}
{{p}}_{1,\text{cm}} = {{p}_{1,\text{lab}} \, M_n \over \sqrt{s}}
\end{equation}
in the lab frame where $M_n$ is at rest and $m$ has three-momentum $p_{1,\text{lab}}$.

\begin{figure*}[!t]
	\subfigure[~scalar mediator, $m_S=0.1~\text{GeV}$.]{\includegraphics[width=8cm]{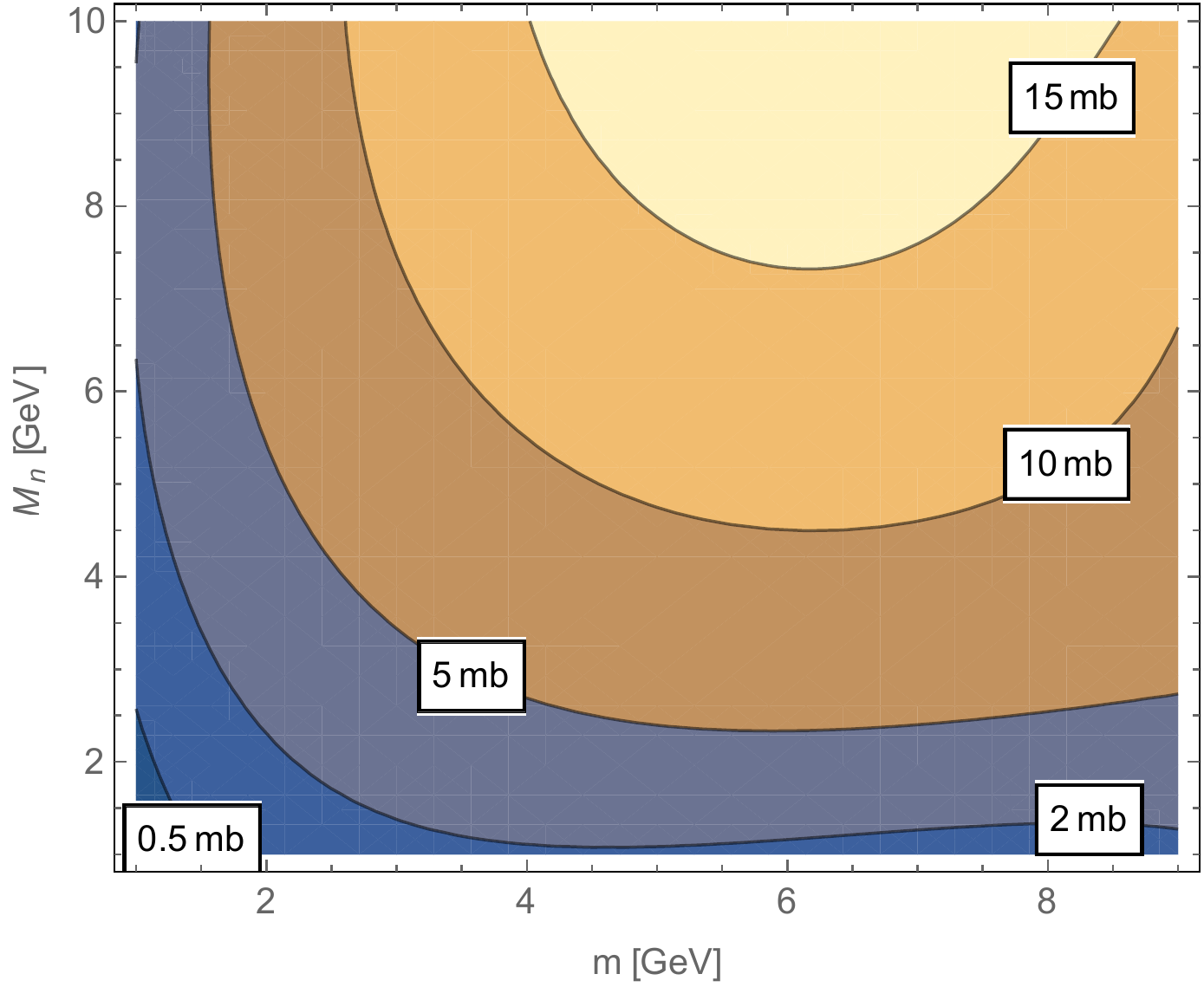}}\hskip1cm
	\subfigure[~scalar mediator, $m_S=100~\text{GeV}$.]{\includegraphics[width=8cm]{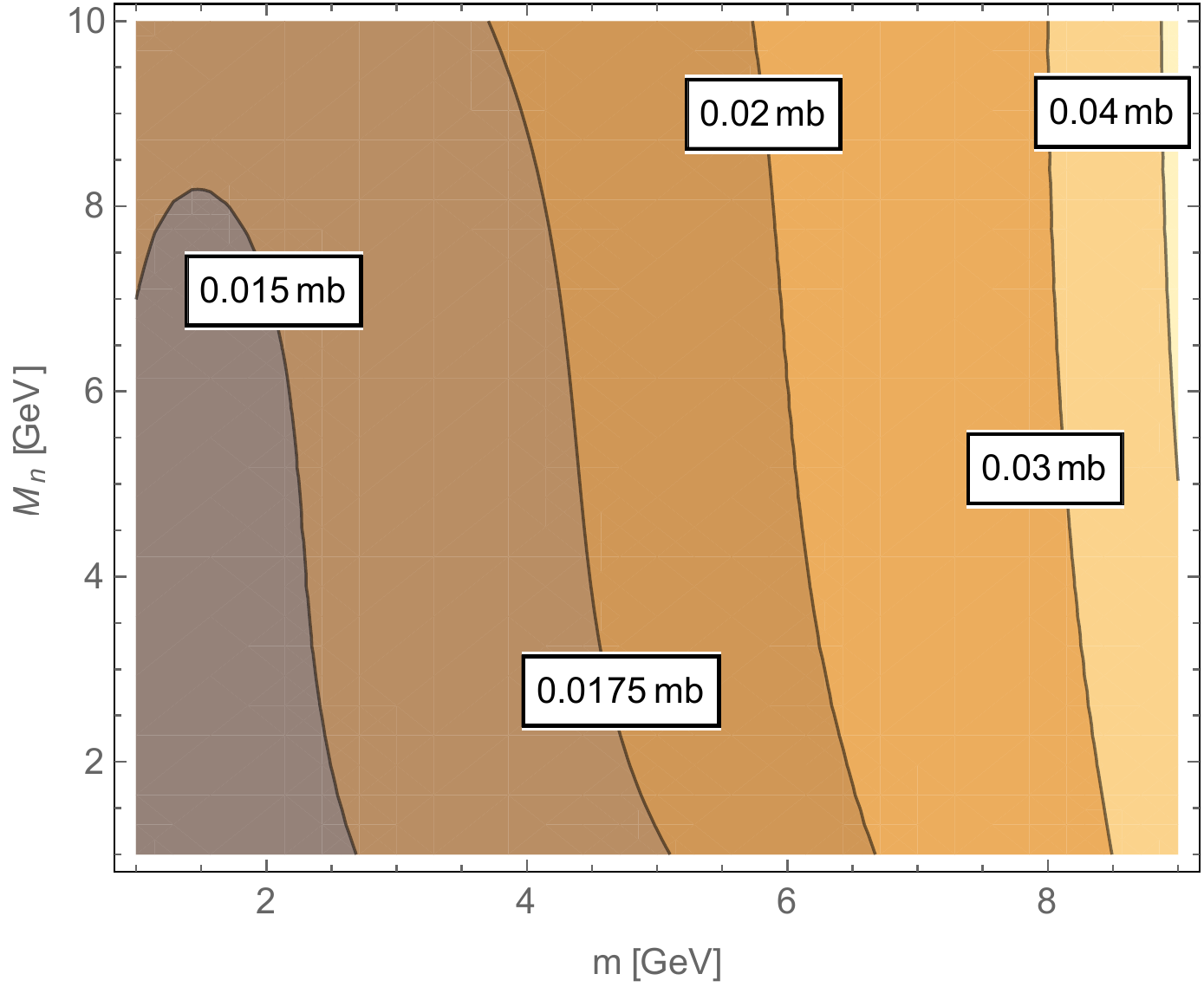}}\\[0.2cm]
	\subfigure[~vector mediator $\sim c_2$, $m_V=0.1~\text{GeV}$.]{\includegraphics[width=8cm]{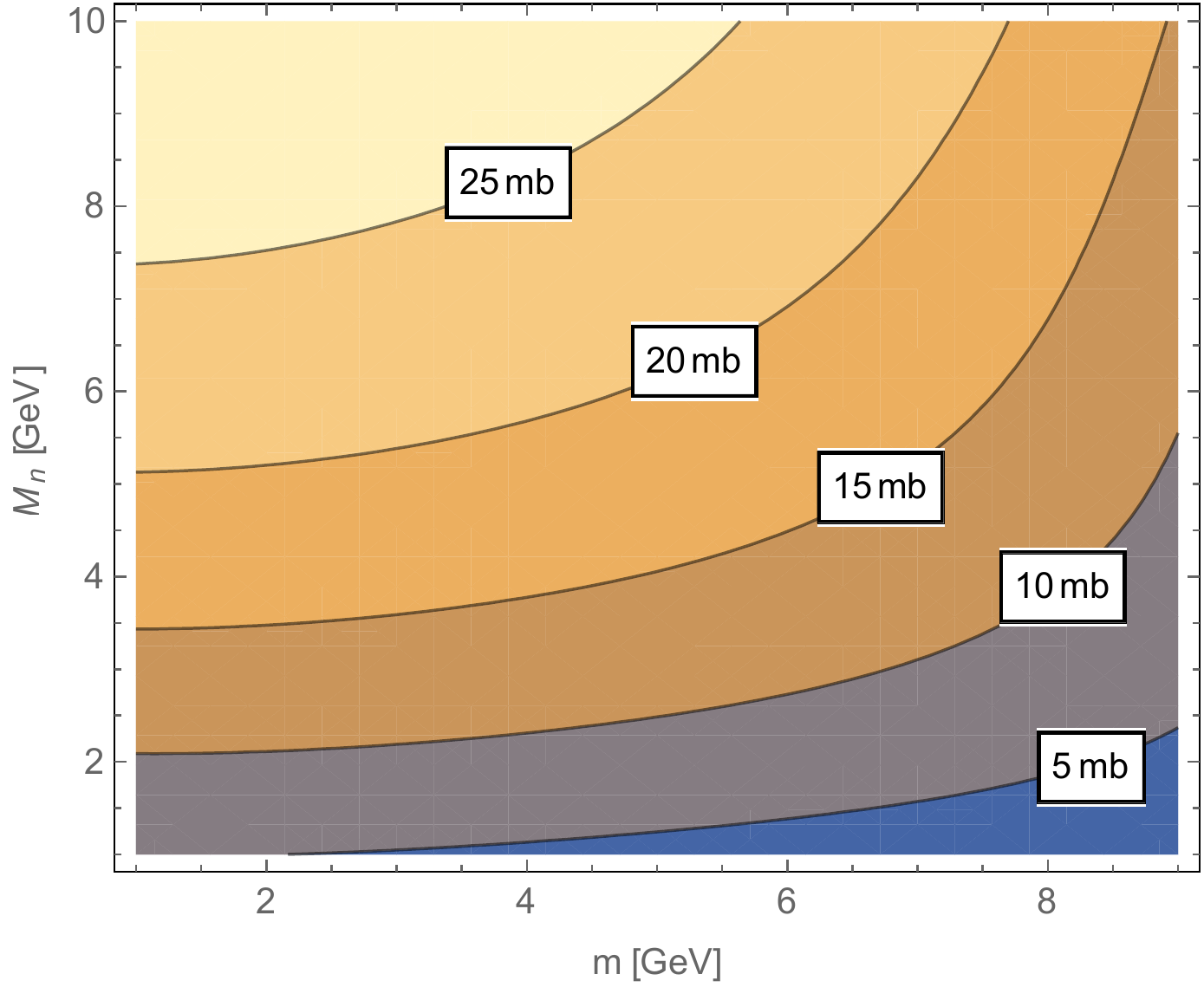}}\hskip1cm
	\subfigure[~vector mediator $\sim c_2$, $m_V=100~\text{GeV}$.]{\includegraphics[width=8cm]{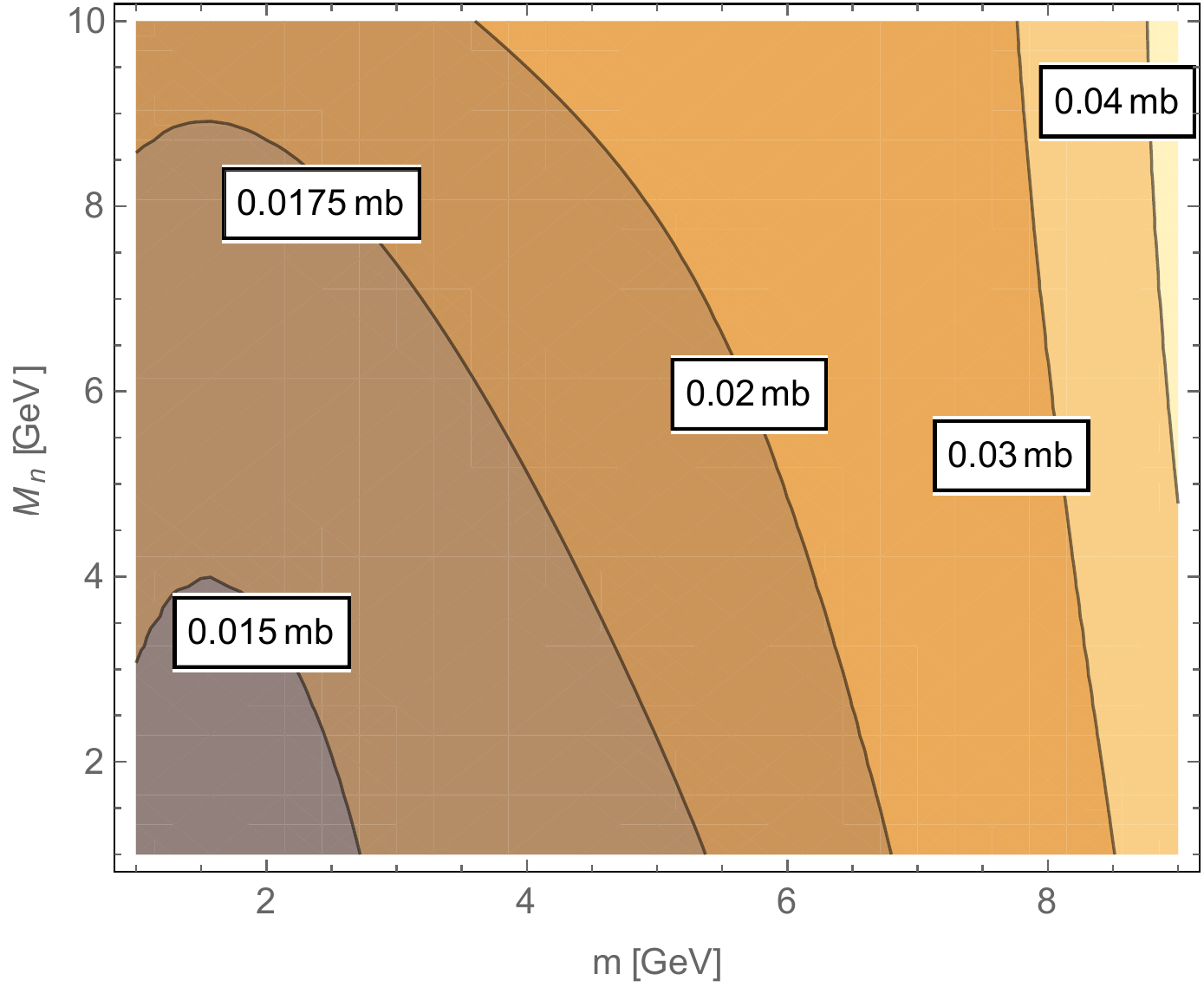}}\\[0.2cm]
	\caption{\label{fig:pressure} Representative total cross section sensitivity contours for the different interactions as defined in Eq.~\eqref{eq:lagrange} and incident beam energy of $E(m)=10~\text{GeV}$ that can be obtained from 
	a sensitivity of $10^{-15}~\text{N}/\sqrt{\text{Hz}}$ at a frequency of $\sim 3~\text{kHz}$ (cf.~Fig.~\ref{fig:Force}).}
\end{figure*}

Projections of future beam facilities suggest that fluxes in the range from $10^{-1}$ to $1$ A are controllable in the beam energy range of $\sim 1~\text{GeV}$~\cite{Shiltsev:2013zma}. Existing neutron beam facilities like PSI, SNS, and LANSCE operate with 1mA in the range of the of $\sim 1~\text{GeV}$, typically using neutrons. The spallation neutron source SINQ operates with a flux of $10^{14}$ neutrons/$\text{cm}^2/\text{s}$. Comparable fluxes at higher energy are more difficult to achieve, however the roadmap of \cite{Shiltsev:2013zma} suggests that high fluxes $\sim 10^{-2}~\text{mA} $ should also be obtainable at future upgraded beam facilities like the FermiLab Booster and NuMI (protons) at energies 10-100 GeV.

For demonstration purposes we assume
\begin{equation}
{\cal{F}}={10^{12} \over {\text{cm}^2~\text{s}}} 
\label{eq:F}
\end{equation}
while the optical thickness of the material 
is of the order of
\begin{equation}
\label{eq:T}
{\cal{T}}= \frac{N_{\mathrm{tot}}}{A} = \frac{d \rho N_A}{m_A} \simeq {1.3 \cdot 10^{22}\over \text{cm}^2}\,, 
\end{equation}
where the constants used describe the absorbers properties are as follows: $N_{\mathrm{tot}}$ its total number of molecules, $A$ is the absorber area, $\rho$ its material density and $d$ its depth. We assumed an absorber of cylindrical shape with $d=0.5$ cm and a total weight of 1 g. $m_A$ is the absorber's material molecular mass of $m_A \simeq 60~\mathrm{g/mol}$, assuming Silicon dioxide $\mathrm{SiO}_2$, and $N_A$ is Avogadro's constant. 
We assume the beam to be focussed on a $10^{-3}~\text{m}$ radius to compute the pressure that can be compared to the intensity curve of Fig.~\ref{fig:intens}. 

Equation~\eqref{eq:cohscatt} shows that uncertainties in the optical thickness can crucially impact the measurement of the pressure. Effectively, the optical thickness plays the role of the luminosity in collider experiments, and it is know that this quantity needs to be precisely know to extract precise theoretical cross sections. We do not include such uncertainties, but note that precise measurements of the optical thickness can be obtained using, e.g., wave-length adjustable lasers~\cite{doi:10.1117/12.472217}.

A given sensitivity threshold of the detector setup (Fig.~\ref{fig:Force}) amounts to an upper limit of $|c_i|$ for given masses of the exchange particles through Eqs.~\eqref{eq:cohscatt} and~\eqref{eq:alpha}. This limit can be interpreted as an upper total cross section limit, see Eq.~\eqref{eq:interaction}. This is shown in Fig.~\ref{fig:pressure}, where we plot the upper cross section limits in the simplified model for our chosen benchmark sensitivity and a range of masses.
The cross section that the pressure measurement is sensitive to shows a significant model-dependence in particular because the pressure measurement highlights the forward and backward scattering kinematics. Therefore, depending on the specific scenario (i.e. nucleons, leptons or even photons as incident beam) as well as the different correlation that is under scrutiny, we can see that the setup discussed in this work could be capable of constraining a range of underlying models

To estimate if gravitational wave detector technology can provide additional insights, beyond previous proton collision experiments, we consider the interaction of an incident beam with $E_\mathrm{beam} = 0.979$ GeV, with a flux of Eq.~\eqref{eq:F} and an absorber as specified in Eq.~\eqref{eq:T}. Assuming a scalar interaction of Eq.~\eqref{eq:lagrange}, i.e. $c_2 = 0$ off a single proton inside the core, and a sensitivity of $10^{-15}~\mathrm{N}/\sqrt{\mathrm{Hz}}$ at $3~\mathrm{kHz}$, we find a sensitivity to a cross section of about $3$ mb within our approximations. This is well beyond the precision of early proton proton experiments~\cite{Patrignani:2016xqp}. As the force can be measured precisely in these setups, decreasing uncertainties in cross section measurements becomes feasible.\footnote{Proton-proton cross sections are well exceeded by proton-nucleus cross sections that are $\gtrsim 100~\text{mb}$ \cite{Ray:1979qk,Carlson:1996ofz,Guzey:2005tk} over a broad range of centre-of-mass energies.}

\begin{figure}[!t]
  \includegraphics[width=7.5cm]{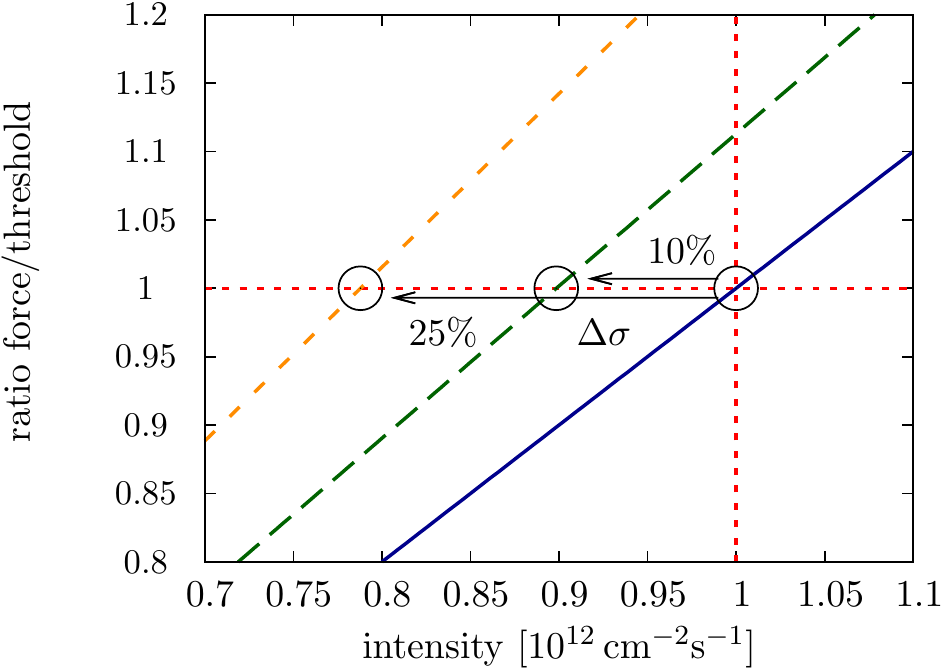}
  \caption{\label{fig:newphys} Different sensitivity thresholds for measured elastic scattering cross sections in the presence of new inelastic interactions that cannot be probed at colliders directly and are based on unitarity arguments. The red lines correspond to a signal hypothesis based on a particular collider cross section measurement. It can be related to the characteristic force that provides an estimate of the beam intensity at which a signal should be detected with the mirror. If a signal is observed at lower intensity, the relative cross section deviation is in one-to-one correspondence with the decreased intensity. This can then be interpreted either as a decreased uncertainty given the model or a force contribution from new scattering processes. We adopt the toy model described above and focus on scattering $m=1~\text{GeV},~M_n=10~\text{GeV}$ with a beam energy of $E\simeq 3~\text{GeV}$.}
\end{figure}

The force measurement can also be used to constrain the presence of new interactions directly or disentangle different contributions through their energy-dependence by varying the beam energy or intensity. In Fig.~\ref{fig:newphys} we demonstrate how additional untagged processes (here assumed to be inelastic scattering $m+M_n\to 0 + M_n$) can be constrained through a force measurement.

\section{Summary}
The progress of gravitational wave detector technology has allowed us to enter an unparalleled regime of precision displacement measurements. It is this progress that
lies at the heart of the direct discovery of gravitational waves. A key question that has been left unaddressed relates to the extent to which this progress can create opportunities
for other areas of physics, possibly beyond the realm of semi-classical approximations. We have addressed this question in this note, demonstrating that the combination of sensitivity
to smallest displacements when paired with modulated particle beams of highest intensity can provide a new avenue to measurements of large interaction cross sections. 
The proposed setup which is based on high-intensity and frequency-controlled beam conditions is key to achieving the best possible cross section constraint. In this sense we provide the first constructive setup of combining particle physics with terrestrial gravitational wave detector technology. Our simplified cross section limits motivate further investigation, not only limited to nucleon interactions where additional effects are likely to influence the ad-hoc sensitivity quoted in this work, but also the consideration of leptons or photons as incident particles. Possible improvements on the measurement side include measuring at different intensities and longer times.

Furthermore, by adding additional material between mirror and incident beam at known expected sensitivity intensities, the setup could be used to provide insights into materials' absorption and transmission properties, but also to provide complementary measurements of nucleon cross sections that are important for, e.g.~dark matter searches. Therefore, if experimentally feasible, the techniques discussed in this work have applications not only in particle and nuclear physics, but also in the field of material sciences or medical applications, e.g. in nuclear therapy where a precise determination of absorption and transmission coefficients of nuclei in biological material is of vital importance for the outcome of the medical procedure.

\acknowledgements
We thank Brian Foster, Lucian Harland-Lang, and Teppei Katori for helpful discussions. 
C.E. is supported by the IPPP Associateship scheme and by the UK Science and Technology Facilities Council (STFC) under grant ST/P000746/1. S.H. is supported by the STFC Grant (CG2016-2020) Ref: ST/N005422/1 and 
the European Research Council (ERC-2012-StG: 307245).


\bibliography{paper.bbl}

\end{document}